\documentclass[
 twocolumn,
 amsmath,amssymb, 
 aps,
 prd,
 letterpaper,
 superscriptaddress,
 showpacs, showkeys,
floatfix,
 nofootinbib,
]{revtex4-2}%

\usepackage[colorlinks,linkcolor=blue,anchorcolor=blue, citecolor=blue,urlcolor=blue,]{hyperref}
\usepackage{graphicx}
\usepackage{color}
\usepackage{xcolor}
\usepackage{float}
\usepackage{makecell}
\usepackage{newtxtext,newtxmath}
\usepackage{fancyhdr}
\usepackage{hyperref}
\usepackage{xcolor}
\allowdisplaybreaks
\usepackage{graphicx}
\usepackage{appendix}
\usepackage{amssymb}
\usepackage{multirow}
\usepackage{booktabs}
\usepackage{float} 
\usepackage{subfigure}
\usepackage{natbib}
\usepackage[marginal]{footmisc}
\usepackage{tikz}
\usepackage{tabularx}
\usepackage{algorithm}
\usepackage{algorithmic}
\usepackage{xcolor}
\usepackage{aas_macros}
\usepackage{enumitem}
\usepackage{ulem}
\usepackage{footnote}

\makeatletter
\usetikzlibrary{shapes.geometric, arrows.meta, positioning}

\begin{document}

\title{Nonlinear Weak Lensing Reconstruction for Galaxy Clusters}
\author{Yuan Shi}
\email{yshi@sjtu.edu.cn}
\affiliation{Department of Astronomy, School of Physics and Astronomy, Shanghai Jiao Tong University, Shanghai, 200240, China}
\affiliation{Key Laboratory for Particle Astrophysics and Cosmology (MOE) /\\ Shanghai Key Laboratory for Particle Physics and Cosmology, China}

\author{Li Cui}
\email{nervous-cl@sjtu.edu.cn} 
\affiliation{Department of Astronomy, School of Physics and Astronomy, Shanghai Jiao Tong University, Shanghai, 200240, China}
\affiliation{Key Laboratory for Particle Astrophysics and Cosmology (MOE) /\\ Shanghai Key Laboratory for Particle Physics and Cosmology, China}

\date{\today}

\begin{abstract}
We present a numerical investigation of nonlinear cluster lens reconstruction using weak lensing mass mapping. Recent advances in imaging and shear estimation have pushed reliable reduced shear measurements closer to cluster cores, making mass reconstruction accessible in the nonlinear regime. However, the Kaiser-Squires based algorithm becomes unstable in cluster cores, where convergence $\kappa$ significantly deviates from zero and the linear approximation breaks down. To address this limitation, we develop a reconstruction framework with two key modifications: applying smooth masks in these regions and using a model-derived analytical solution as the initial guess, rather than assuming $\kappa = 0$. We validate our framework using simulated cluster lensing data with known mass distributions, incorporating realistic masks that arise from limitations in reduced shear measurements.
We show that in the absence of shape noise, our framework yields high-fidelity mass reconstruction in regions with large reduced shear, with the best-performing method achieving residuals below $0.02 \sigma$ in the unmasked regions. We further discover that iterative reconstruction methods can diverge beyond an optimal iteration count. Our modifications effectively suppress this instability, pushing mass reconstruction to higher accuracy and
stability in the nonlinear regime.
\end{abstract}

\maketitle

\section{Introduction} \label{sec:intro}
Clusters of galaxies are powerful tools for understanding structure formation and constraining cosmological models \cite{Mass_07, Witt, Allen_11, Pratt_19}. However, fully exploiting their potential requires an accurate reconstruction of cluster mass distributions. Among various methods for measuring cluster masses, weak gravitational lensing (WL) is widely recognized as a crucial probe, as it directly measures the projected mass along the line of sight, independent of the cluster’s dynamical state \cite{Hoek08, Umetsu14, Umetsu20, kneib_cl, huanyuan_10}. 

Recent advances in Stage~III and IV imaging surveys have significantly improved source densities and image quality, enabling reduced shear measurements closer to cluster cores \cite[e.g.,][]{Abbott2016, Aihara2018, Laureijs2011, Euclid2022, Ivezi2020, Spergel2015, Yao2024CSST}. In the central regions of clusters, particularly those massive enough to produce strong lensing phenomena, the reduced shear $g$ is sufficiently large that the weak lensing approximation no longer holds. When the reduced shear exceeds $\sim0.2$, systematic biases arise in shape measurements due to nonlinear effects \cite{Becker11, Hern_20, sun_25}, while regions with $|g| \gtrsim 0.5$ yield no reliable measurements and are therefore masked in weak lensing analyses \cite{Harvey24}. Together with masks from saturated stars and other observational artifacts, these masked regions complicate accurate mass reconstruction \cite{Pires2020}. 

So far, weak lensing mass reconstruction of galaxy clusters has been traditionally classified into two categories: parametric and non-parametric methods \cite{Wright_00, Mandel_10,S&K, Lanusse_16}. More recently, machine learning techniques have emerged as a promising data-driven approach, enabling weak-lensing reconstructions without imposing explicit parametric mass models \cite{Shira_19, Hong_21, Cha_25}. Among non-parametric approaches, iterative Kaiser-Squires (KS) algorithms are widely used \cite{S&S}. However, their performance is affected by the presence of masked regions, which are typically treated as zero-valued data in the reconstruction. Since the mapping from shear to convergence is intrinsically non-local, instabilities originating in masked regions propagate throughout the field, introducing biases in the recovered convergence map.  


In this work, we use simulated reduced shear maps constructed from two configurations: idealized toy models and cluster mass profiles derived from JWST strong lensing analyses. This setup allows us to address three critical issues. 
First, we consider masks arising from unreliable reduced shear measurements and quantify the resulting bias. Second, we examine how reconstruction accuracy and stability vary with iteration count in the presence of masks. Finally, we explore the systematic limits of mass reconstruction in increasingly nonlinear regimes by varying the threshold of available reduced shear data.
This investigation directly probes how far a weak lensing mass reconstruction can reliably reach into the cluster core, a fundamental question for current and next-generation surveys that seek to maximize the reconstruction area in cluster mass maps.


The paper is structured as follows. In Sec. $2$, we detail the mass reconstruction methodologies developed in this study. In Sec. $3$, we present the simulation setup and the resulting mock data, and evaluate the performance of different methods under various conditions. Sec. $4$ we summarize our findings and discuss prospects for future studies. 
Throughout this paper, we adopt a standard flat $\Lambda$CDM cosmology with $\Omega_{\mathrm{m}} = 0.3$, $\Omega_{\Lambda} = 0.7$, and $h = 0.7$.

\section{Methods} \label{sec:methods}
In this section, we begin with a review of previous mass mapping approaches, followed by the modifications implemented in this work.
\subsection{KS and AKRA} \label{subsec:ks_akra}

The convergence $\kappa$ and shear components $\gamma_{1,2}$ at position $\boldsymbol{\theta} = (\theta_1, \theta_2)$ are derived from lensing potential $\psi$ as \cite{Sch92, Bar_01}
\begin{equation}
    \kappa = \frac{1}{2}(\partial_1^2 + \partial_2^2) \psi, \quad \gamma_1 = \frac{1}{2}(\partial_1^2 - \partial_2^2) \psi, \quad \gamma_2 =\partial_1\partial_2 \psi,
\end{equation}
In Fourier space, derivatives become multiplications: $\partial_i \rightarrow i\ell_i$, where $\boldsymbol{\ell} = (\ell_1, \ell_2) = |\boldsymbol{\ell}|(\text{cos} \phi_\ell, \text{sin}\phi_{\ell})$ is the wavevector conjugate to $\boldsymbol{\theta}$. This yields \cite{Ks93}
\begin{equation}
    \begin{bmatrix} 
    \tilde{\gamma_1}(\boldsymbol{\ell}) \\ \tilde{\gamma_2}(\boldsymbol{\ell}) \end{bmatrix} = \begin{bmatrix} \cos(2\phi_\ell)  \\ \sin(2\phi_\ell) \end{bmatrix} \tilde{\kappa}(\boldsymbol{\ell}),
    \label{eq:ks}
\end{equation}
Inverting this relation, we obtain
\begin{equation}
    \tilde{\kappa}(\vec{\ell}) = \tilde{\gamma}_{1}(\vec{\ell}) \cos(2\phi_{\ell}) + \tilde{\gamma}_{2}(\vec{\ell}) \sin(2\phi_{\ell}).
    \label{eq:ks_fourier}
\end{equation}
Transforming back to real space, multiplication becomes a convolution \cite{SL_2022}
\begin{equation}
    \kappa(\boldsymbol{\theta}) - \kappa_0 = \frac{1}{\pi} \int d^2\theta' \Re[\mathcal{D}^*(\boldsymbol{\theta} - \boldsymbol{\theta}') \gamma(\boldsymbol{\theta}')],
\end{equation}
with kernel
\begin{equation}
    \mathcal{D}(\boldsymbol{\theta}) \equiv 
    \frac{\theta_2^2-\theta_1^2 -2\text{i}\theta_1\theta_2}{|\boldsymbol{\theta}|^4},
\end{equation}

However, in practice the observables are axis ratio R and angle of the major axis relative to the x-axis $\varphi_0$ \cite{Nick95}, from which the reduced shear $g = \gamma / (1-\kappa)$ is estimated.

In the weak lensing regime where $\kappa \ll 1$,the approximation $g \simeq \gamma$ holds and the above equation can be applied directly. This approximation breaks down in massive clusters where $\kappa$ is significant. Since only the reduced shear g is accessible from observations, the inversion is reformulated as \cite{S&S}
\begin{equation}
    \kappa(\boldsymbol{\theta}) -\kappa_0 = \frac{1}{\pi} \int d^2\theta' \Re[\mathcal{D}^*(\boldsymbol{\theta} - \boldsymbol{\theta}') g(\boldsymbol{\theta}')(1-\kappa(\boldsymbol{\theta}'))].
\end{equation}
Thus, the convergence map is obtained using an iterative scheme, initialized with $\kappa = 0$ and refined through iterative updates until convergence. As discussed earlier, this approach introduces bias in the presence of masks. Therefore, we also apply the Accurate Kappa Reconstruction Algorithm (AKRA) \cite{shi2024, shi2025}, to fully investigate the three issues outlined in the introduction.

Here we briefly review the theoretical basis of AKRA. Mathematically, the effect of masking can be described by introducing a binary mask function $m(\boldsymbol{\theta})$, taking the value 1 in unmasked regions and 0 in masked regions. The masked shear is then expressed as
\begin{equation}
    \gamma_i^{\mathrm{m}}(\boldsymbol{\theta}) = m(\boldsymbol{\theta}) \, \gamma_i(\boldsymbol{\theta}).
    \label{eq
:masked_shear}
\end{equation}
Since multiplication in real space on pixelized maps corresponds to discrete convolution in Fourier space, this convolution can in turn be expressed as matrix multiplication. Therefore, we introduce the matrix $\mathbf{M}$ to represent the convolution with the mask. In Fourier space, the relation between masked shear and convergence turns into
\begin{equation}
    \begin{bmatrix} \tilde{\gamma}_1^{\mathrm{m}}(\boldsymbol{\ell}) \\ \tilde{\gamma}_2^{\mathrm{m}}(\boldsymbol{\ell}) \end{bmatrix} = \begin{bmatrix} \cos(2\phi_\ell) \, \mathbf{M} \\ \sin(2\phi_\ell) \, \mathbf{M} \end{bmatrix} \tilde{\kappa}(\boldsymbol{\ell}),
    \label{eq:masked_fourier}
\end{equation}

To obtain a complete theoretical description of the mass reconstruction procedure, we account for observational uncertainties by introducing a noise term in matrix form $\boldsymbol{n}$, and reformulate Eq.~\eqref{eq:masked_fourier} as a linear system in Fourier space
\begin{equation}
\boldsymbol{\gamma^{\mathrm{m}}} = \mathbf{A}\boldsymbol{\kappa} + \boldsymbol{n},
\label{eq:linear_system}
\end{equation}
where $\boldsymbol{\gamma^{\mathrm{m}}} = [\tilde{\gamma}_1^{\mathrm{m}}, \tilde{\gamma}_2^{\mathrm{m}}]^{\mathrm{T}}$ is the masked shear data vector, $\boldsymbol{\kappa}$ is the convergence vector in Fourier space, $\mathbf{N} \equiv \langle \boldsymbol{n}\,\boldsymbol{n}^{\mathrm{T}} \rangle$ is the noise covariance matrix and $\mathbf{R} = \lambda \mathbf{I}$
(with $\lambda \sim 10^{-3}$) is a regularization term included for numerical stability
\footnote{The regularization parameter $\lambda$ controls the trade-off between fidelity to the data and numerical stability of the inversion. Its physical role can be understood from the eigenvalue structure of $\mathbf{A}^{\rm T}\mathbf{N}^{-1}\mathbf{A}$. In the idealized case without masks, $\mathbf{M} \to \mathbf{I}$, and the normal matrix reduces to $\mathbf{A}^{\rm T}\mathbf{A} = \mathrm{diag}(\cos^2 2\phi_{\ell}) + \mathrm{diag}(\sin^2 2\phi_{\ell}) = \mathbf{I}$, so all eigenvalues are unity. In the presence of a survey mask, mode coupling redistributes eigenvalue weight, driving a subset of eigenvalues toward zero. The addition of $\lambda\mathbf{I}$ raises every eigenvalue by $\lambda$, bounding the condition number of the system and stabilizing the inversion.
The systematic bias introduced by regularization can be estimated analytically. For a mode with true eigenvalue $\mu$, the regularized estimator recovers a fraction $\mu/(\mu + \lambda)$ of the true signal. Since the vast majority of eigenvalues satisfy $\mu \gg \lambda$ for $\lambda = 10^{-3}$ (see Figs.~18 and 19 of Ref.~\cite{shi2024} for the full eigenvalue distribution), the fractional suppression is $\lesssim \lambda/(1+\lambda) \approx 0.1\%$.}
.
The matrix $\mathbf{A}$ encodes both the lensing response and the mask convolution
\begin{equation}
    \mathbf{A} = \begin{bmatrix} \cos(2\phi_\ell) \, \mathbf{M} \\ \sin(2\phi_\ell) \, \mathbf{M} \end{bmatrix},
    \label{eq:A_matrix}
\end{equation}
In the absence of masks, $\mathbf{M}$ reduces to the identity matrix and AKRA takes the same form as Eq.~\eqref{eq:ks}. The detailed algorithm for computing the mask matrix $\mathbf{M}$ is described in Appendix~\ref{app:mask}.

\begin{table*}[ht]
\centering
\caption{Comparison of standard and modified methods.}
\label{tab:comparison}
\begin{tabular}{lccccc}
\hline
\hline
Method & K1 & K2 & A1 & A2 & A3 \\
\hline
Initial guess & $\kappa^{(0)} = 0$ & $\kappa^{(0)} = \kappa^{\mathrm{model}}$ & $\kappa^{(0)} = 0$ & $\kappa^{(0)} = \kappa^{\mathrm{model}} $ & $\kappa^{(0)} = \kappa^{\mathrm{model}}$ \\
Mask type & Binary & Binary & Binary & Binary & Smooth \\
Mass mapping & KS & KS & AKRA & AKRA & AKRA \\
\hline
\end{tabular}
\end{table*}

The convergence is then obtained by applying the AKRA estimator to the corrected shear
\begin{equation}
    \hat{\boldsymbol{\kappa}} = \left( \mathbf{A}^{\mathrm{T}} \mathbf{N}^{-1} \mathbf{A} + \mathbf{R} \right)^{-1} \mathbf{A}^{\mathrm{T}} \mathbf{N}^{-1} \boldsymbol{\gamma^{\mathrm{m}}},
\label{eq:kappa_akra}
\end{equation}
To apply this framework to real data,  we employ the same iterative approach described earlier, substituting $\gamma^{\mathrm{m}} = g^{\mathrm{m}}(1-\kappa)$ and updating $\kappa$ until convergence. The iterative form of AKRA is then given by
\begin{equation}
    \hat{\boldsymbol{\kappa}}^{(i)} = \left( \mathbf{A}^{\mathrm{T}} \mathbf{N}^{-1} \mathbf{A} + \mathbf{R} \right)^{-1} \mathbf{A}^{\mathrm{T}} \mathbf{N}^{-1}
    \boldsymbol{\mathcal{F}}\left[{g^{\mathrm{m}}}(1- {\kappa}^{(i-1)})\right].
\label{eq:it_akra}
\end{equation}
where $\boldsymbol{\mathcal{F}}$ denotes the Fourier transform, the superscript i denotes the iteration number. The final convergence map is then obtained by transforming $\hat{\boldsymbol{\kappa}}$ back to real space after reaching convergence.


\subsection{Modifications} \label{subsec:modifications}
We initialize the mass reconstruction with a model-based approximation rather than assuming $\kappa^{(0)}=0$ across the field. The convergence can deviate significantly from zero in clusters, especially in the dense core, making this assumption a poor initial estimate. We therefore apply the Singular Isothermal Sphere (SIS) model to construct the initial guess. Since $|\gamma| = \kappa$ for SIS, the reduced shear simplifies to 
\begin{equation}
    g = \frac{\kappa}{1-\kappa}
\end{equation}
from which we obtain the intial $\kappa^{(0)}$
\begin{equation}
    \kappa^{(0)} = \frac{|g|}{1+|g|}
\end{equation}

Having addressed the initialization, we now turn to the functional form of the mask. A key limitation of the binary mask function $m(\boldsymbol{\theta})$ is its inherent discontinuity, which can cause instabilities in the reconstruction. To address this limitation, we generalize $m(\boldsymbol{\theta})$ as a smooth transition function
\begin{equation}
    m(\boldsymbol{\theta}) = 
    \begin{cases}
        1, & g(\boldsymbol{\theta}) < g_* \\[4pt]
        (1 + \sin\xi(\boldsymbol{\theta}))/2, & g_* \leq g(\boldsymbol{\theta}) \leq g_{\rm th} \\[4pt]
        0, & g(\boldsymbol{\theta}) > g_{\rm th}
    \end{cases}
    \label{eq:smooth_mask}
\end{equation}
where $\xi(\boldsymbol{\theta}) = [(g_{\rm th} - g(\boldsymbol{\theta}) )/(g_{\rm th} - g_*) - \frac{1}{2}]\pi$. This mask function is defined by two characteristic values: $g_*$ and $g_{\rm{th}}$. The former sets the value at which the weight begins to decrease from unity, while the latter defines the threshold beyond which measurements are fully masked. 
In practice, $g_{\rm th}$ is set by the observational catalog, marking the cutoff beyond which reduced shear measurements are not provided.  $g_*$ denotes the value at which the uncertainty in reduced shear measurements increases significantly. In the inner regions of clusters, light from foreground cluster galaxies and diffuse intracluster light reduces the effective number density of background sources, which in turn increases shape noise and degrades shear measurement precision. This smooth transition form gradually downweights measurements in this regime. In the limit $g_* \to g_{\rm th}$, the mask function reduces to the binary form.

The combination of different initializations and mask functions yields five configurations, which we summarize in Table~\ref{tab:comparison}.
\\

\begin{figure*}[ht]
    \centering
    \includegraphics[width=0.86\textwidth]{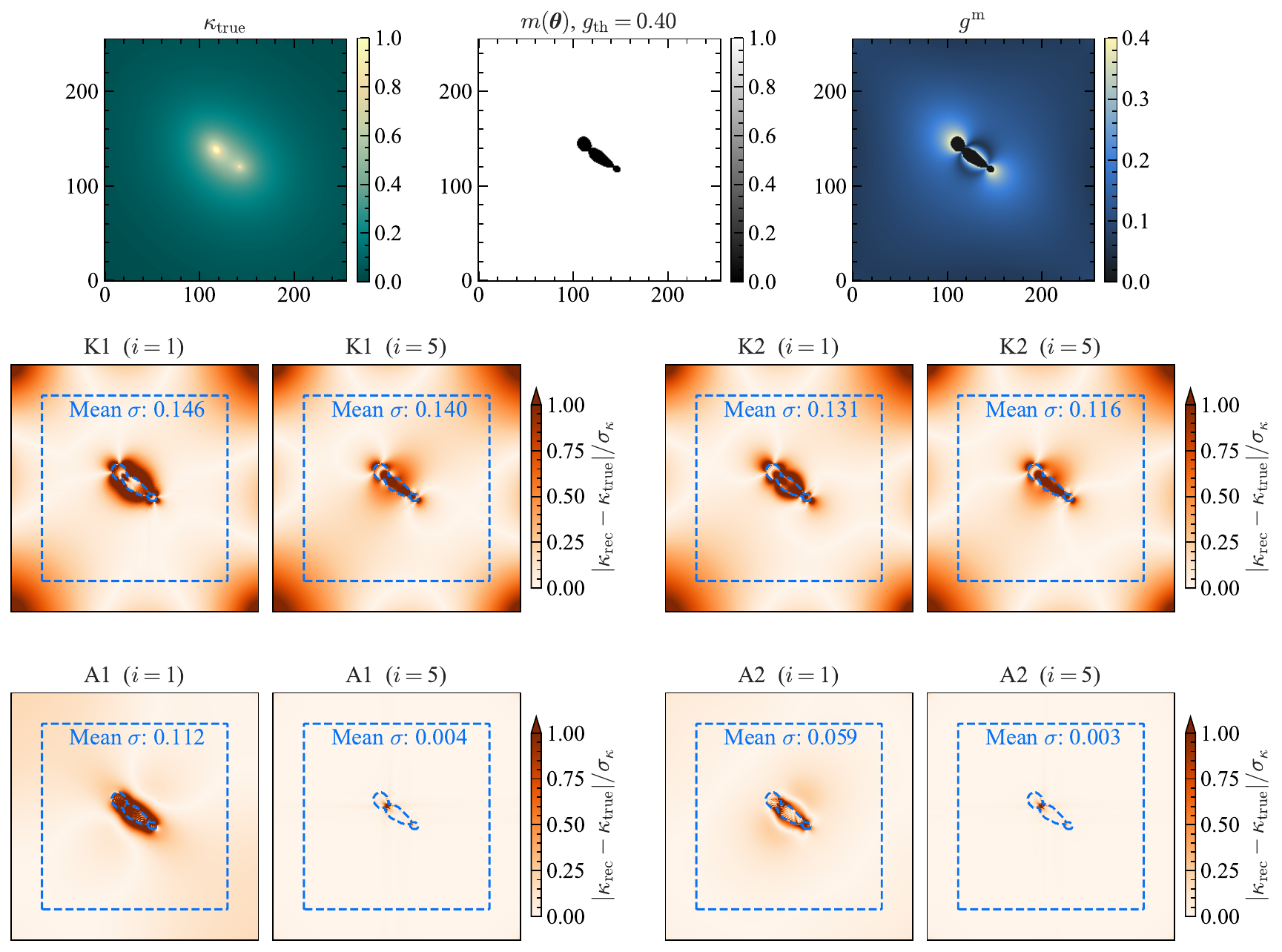}
    \caption{Comparison of reconstructed convergence maps for the toy model. \textit{Top panels}: True convergence $\kappa_{\rm true}$, the binary mask with $g_{\rm th} = 0.4$ and masked reduced shear $g^{\rm m}$. \textit{Middle panels}: Normalized residual $\sigma$ (Eq.~\ref{eq:sigma}) for KS-based methods K1 and K2 at first and fifth iterations. \textit{Lower panels}: Same as middle panels but for AKRA-based methods A1 and A2. The blue dashed box indicates the region where the mean $\sigma$ is computed. The mean $\sigma$ values are shown in each panel. The inner blue dashed contour outlines the masked region.}
    \label{fig:kappa_toy}
\end{figure*}

\section{Numerical Tests}
\subsection{Mock catalogs}
To assess the performance of different mass reconstruction methods, we generate two mock clusters with known mass distributions. The first is an idealized toy model constructed using the publicly available lensing software \texttt{GLAFIC} \cite{Oguri10}. It consists of two NFW halos \cite{Na96} at $z=0.3$, with virial masses $M_1 = 4.3 \times 10^{14} \, h^{-1} \, M_{\odot}$ and $M_2 = 3.2 \times 10^{14} \, h^{-1} \, M_{\odot}$, separated by a projected distance of $\sim 140 \, h^{-1} \, {\rm kpc}$. An external shear term is also included to account for line-of-sight tidal perturbations. The field of view is $6\times6~{\rm arcmin}^2$.

The second is based on the strong lensing mass model of Abell 2744 ($z=0.308$) from \citet{2023b}, hereafter referred to as B23 model, and we adopt a field of $10 \times 10~{\rm arcmin}^2$. Abell 2744 is a massive, dynamically disturbed cluster with multiple mass peaks and complex internal structure \cite{Merten_11, Owe11, Med_16, Cha_24}. Its complex structure and dense cores make it well suited for testing the limits of weak lensing mass mapping. For both models, we derive the shear field from the convergence map and compute the corresponding reduced shear, which is then used as input for the reconstruction. The data are assumed to be noise-free and sampled on a $256\times256$ grid, which enables an examination of systematic effects in the absence of statistical noise.


\subsection{Mass mapping results} \label{sec:results}

To quantify the accuracy of the reconstructed convergence maps, we define a normalized residual metric
\begin{equation}
    \sigma(\boldsymbol{\theta}) = \frac{|\kappa_{\rm rec}(\boldsymbol{\theta}) - \kappa_{\rm true}(\boldsymbol{\theta})|}{\sigma_\kappa}.
    \label{eq:sigma}
\end{equation}
where $\kappa_{\rm rec}$ and $\kappa_{\rm true}$ denote the reconstructed and true convergence, and $\sigma_\kappa$ is the standard deviation calculated from true convergence field.

\begin{figure*}[ht]
    \centering
    \includegraphics[width=0.86\textwidth]{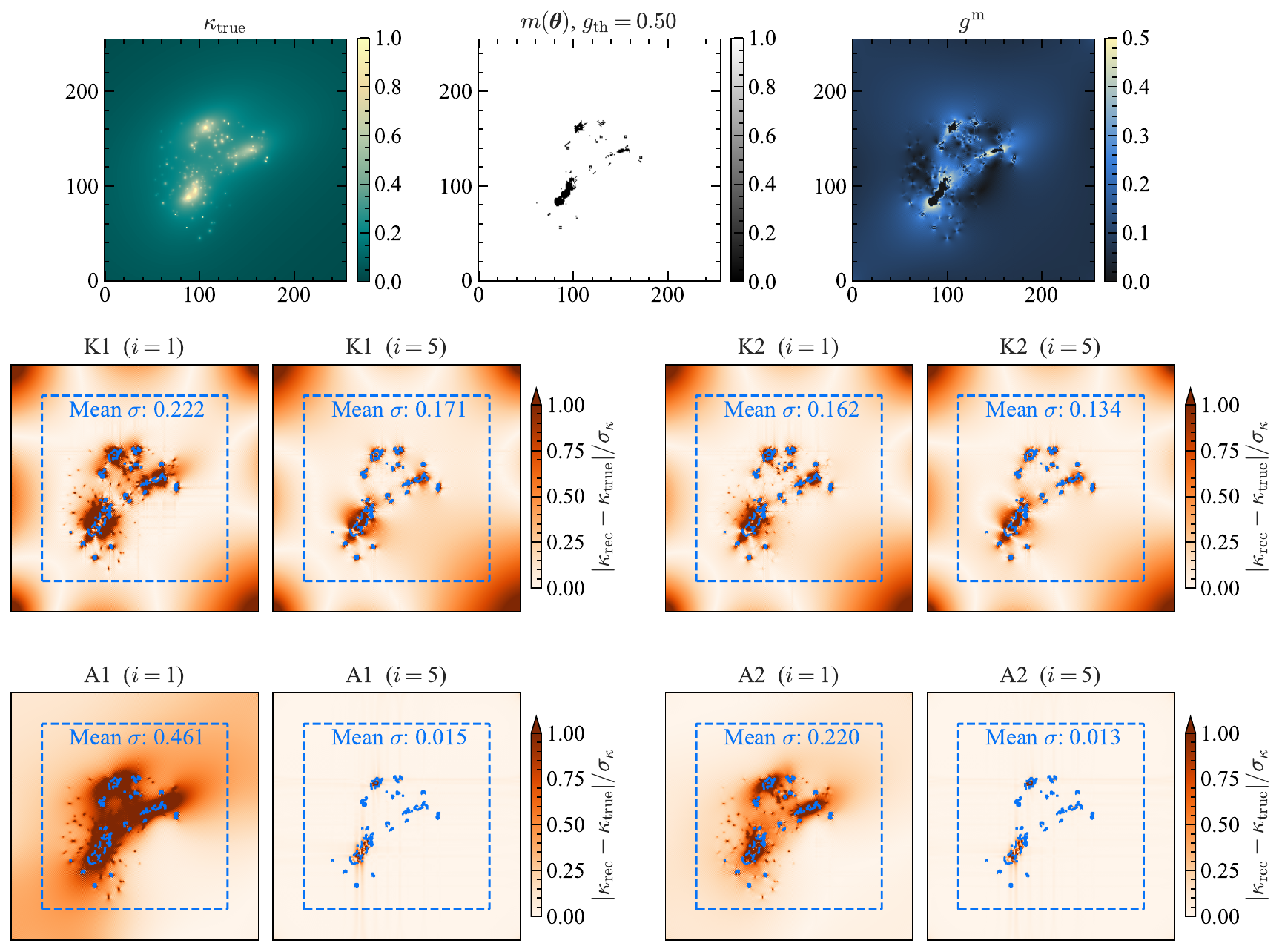}
    \caption{Reconstructed convergence maps for the B23 model. \textit{Top panels}: True convergence map, binary mask with $g_{\rm th} = 0.5$, and the masked reduced shear $g^{\rm m}$. \textit{Middle and lower panels}: Normalized residuals for KS-based and AKRA-based methods, respectively.}
    \label{fig:kappa_abell}
\end{figure*}

We evaluate the reconstruction accuracy in the unmasked region where $g(\boldsymbol{\theta}) < g_{\rm th}$. 
Since the Fourier transform assumes periodic boundary condition, the $256 \times 256$ field is zero-padded to $512 \times 512$. This padding introduces discontinuities that cause edge artifacts in iterative KS methods. We therefore exclude 32 pixels from each side along both spatial directions, leaving a central $192 \times 192$ grid for computing summary statistics. This region is shown as the blue dashed box in Fig.~\ref{fig:kappa_toy} and Fig.~\ref{fig:kappa_abell} , where the mean residual $\langle \sigma \rangle$ is computed to quantify the reconstruction accuracy.

\subsubsection{Toy model}

We begin by quantifying the systematic bias introduced by a binary-valued mask function, as commonly implemented in weak lensing mass reconstruction. To investigate this, we generate a mock cluster and apply a mask to pixels where the reduced shear exceeds $g_{\rm th}=0.4$, a regime where shape measurements are typically considered unreliable and are therefore excluded from weak lensing analyses. 

We apply four reconstruction methods to this mock data: K1, K2, A1 and A2 (see Table~\ref{tab:comparison}), and present the results in Fig.~\ref{fig:kappa_toy}. 
Our results show that in the unmasked regions, the KS-based methods converge to a mean residual of $\langle \sigma \rangle \approx 0.12$, whereas the AKRA-based methods achieve an order of magnitude lower residual, with $\langle \sigma \rangle \approx 0.003$. This indicates that the AKRA-based approach substantially reduces this systematic bias. Furthermore, the improvement from K1 to K2 and from A1 to A2 illustrates the effect of our modified initial guess, leading to lower variance and enhanced numerical stability.

We also note that during the iterative procedure, the bias initially decreases  but increases beyond a certain number of iterations. This is particularly relevant when mass reconstruction is applied to real observations, where the stopping criterion is often set to a fixed number, which may exceed the optimal point and introduce additional bias. This behavior is shown in Appendix~\ref{app:iterations}.

\begin{figure*}[ht]
    \centering
    \includegraphics[width=0.99\textwidth]{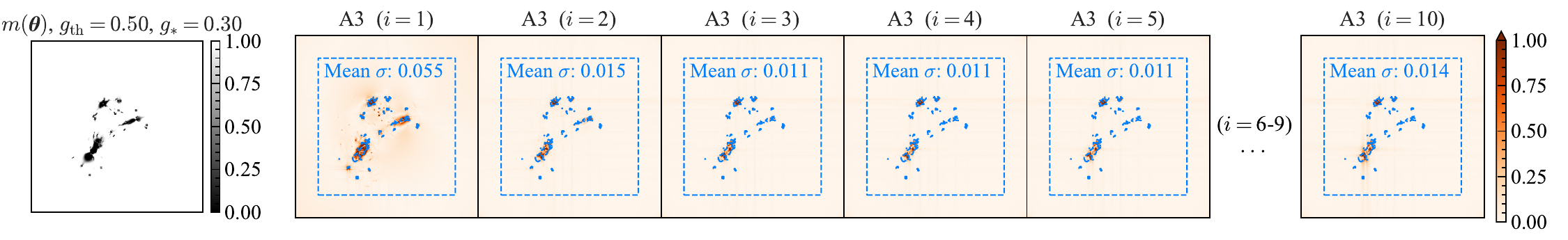}\\
    \includegraphics[width=0.99\textwidth]{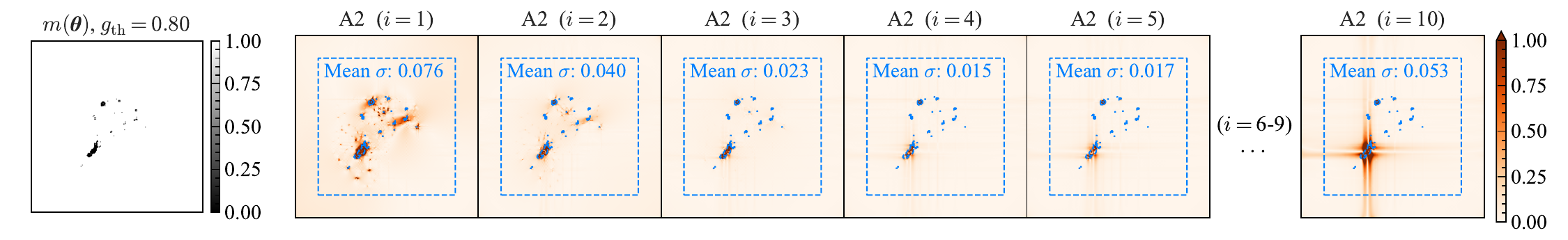}\\
    \includegraphics[width=0.99\textwidth]{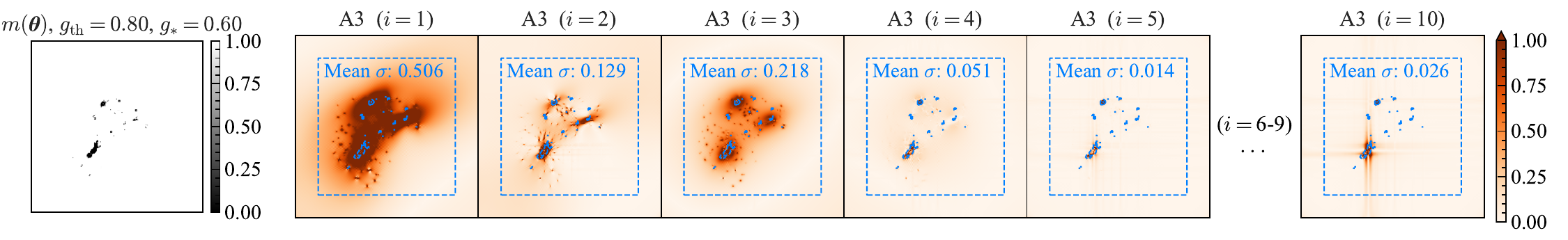}\\
    \caption{Iterative reconstruction results under different masking schemes. \textit{Top panels:} A3 with smooth mask ($g_{\rm th} = 0.5$, $g_* = 0.3$) and normalized residuals. \textit{Middle panels:} A2 with binary mask ($g_{\rm th} = 0.8$) and normalized residuals. \textit{Bottom panels:} A3 with smooth mask ($g_{\rm th} = 0.8$, $g_* = 0.6$) and normalized residuals.}
    \label{fig:A23_iter}
\end{figure*}

\subsubsection{B23 model}
We now turn to a more realistic case based on the B23 model. The convergence map of the B23 model provides access to the full cluster mass distribution, allowing us to generate reduced shear fields for arbitrary choices of $g_{\rm th}$. This capability makes it well suited for testing how far inward the iterative mass mapping method remains reliable. Guided by the shear catalog from \citet{Harvey24}, we start by adopting a binary mask with $g_{\rm th} = 0.5$. Applying the same reconstruction methods as described above, we find results consistent with the toy model: the KS-based methods produce similar mean residuals $(\langle \sigma \rangle \approx 0.13)$, while the AKRA-based methods lower this bias by an order of magnitude, as illustrated in Fig.~\ref{fig:kappa_abell}. 

We next investigate the intrinsic limitations of mass-mapping methods, taking masked reduced shear data as input. The specific question is whether reconstruction methods can recover reliable solutions in unmasked regions under different mask thresholds. We address this by applying A2 and A3 to the reconstruction. Since A2 yields the lowest residuals in the earlier comparison, it is adopted as the reference for assessing A3. We then examine how their performance varies with mask thresholds, as shown in Fig.~\ref{fig:A23_iter}.

In the following, we outline the reconstruction results under different masking schemes. When comparing the two masking schemes under the same $g_{\rm th}$, the smooth mask initially produces larger residuals than the binary mask during the first few iterations, due to its information loss in the region where $g(\boldsymbol{\theta}) \in (g_* ,g_{\rm th})$. In the later stages, the modification with a smooth mask provides greater numerical stability and yields smaller residuals upon convergence. Overall, our methods achieve convergence with low residuals even at $g_{\rm th} = 0.8$, where the input reduced shear exhibits sharp edges.

These results probe the intrinsic boundary of weak lensing mass reconstruction in the nonlinear regime. As $g_{\rm th}$ increases, the reconstruction extends deeper into the cluster core, where the reduced shear is large. Both A2 and A3 achieve low residuals up to $g_{\rm th} = 0.8$, demonstrating that the iterative method remains reliable even in this regime. This indicates that the reconstruction boundary is determined by data quality, rather than by the method itself.

Moreover, A3 exhibits smaller variation in residuals across iterations compared to A2. When applied to real data, the true convergence field is unknown, making the optimal stopping iteration difficult to determine. The stability of A3 across 
iterations makes it less sensitive to this choice. This advantage becomes increasingly relevant for next-generation surveys, where improved measurement precision will push $g_{\rm th}$ further into the cluster core, producing a sharper transition at the mask boundary than in current data. These properties enable A3 to provide reliable reconstructions under such conditions.


\section{Conclusion and discussion}

We have compared five methods for weak-lensing mass mapping on simulated galaxy clusters, including the existing approaches (K1, A1) and our modified versions (K2, A2, A3), which we summarize below in terms of their practical implications. 

\begin{enumerate}
    \item Within KS-based frameworks, K2 can be widely incorporated into existing iterative algorithms to improve numerical stability and achieve lower residuals.
    
    \item For analyses requiring both computational efficiency and reconstruction accuracy, A2 provides an optimal balance and is well suited for most applications.
    
    \item For high-precision or next-generation applications, A3 achieves the highest stability and reconstruction accuracy, with its advantages becoming particularly pronounced when reduced shear measurements are available in strongly nonlinear regimes.
\end{enumerate}

In this work, we adopt relatively simple and reasonable forms for the initial guess and smooth mask function, but determining their optimal forms remains unresolved. We also note that residuals begin to increase after a certain number of iterations in simulations, suggesting that careful consideration is required when determining the optimal stopping point for real observational data. 

We present a framework that improves iterative weak lensing reconstruction in high reduced shear regimes, reliably recovering the convergence over a broader radial range. This advancement enables more robust joint analyses with complementary mass tracers, tighter dark matter–baryon constraints, and improved substructure identification.
In future work, we will apply these methods to real observational data to further validate their performance under realistic conditions. All codes will be made publicly available upon publication. 

\section*{Acknowledgements}
We thank the anonymous referee for valuable suggestions and helpful comments.
We thank Pengjie Zhang for useful suggestions. 
We sincerely thank Carlo Giocoli, Tian-Xiang Mao for helpful discussions on the simulation setup.
YS acknowledges the support from NSFC Grant No. 12503004.
This work is also supported by the National Key R\&D Program of China (2023YFA1607800, 2023YFA1607801), and the Fundamental Research Funds for the Central Universities. 
This work made use of the Gravity Supercomputer at the Department of Astronomy, Shanghai Jiao Tong University.
We also acknowledge the use of the following software packages: \texttt{GLAFIC} \cite{Oguri10, Oguri_21}, \texttt{Lenstool} \cite{Kneib2011, Jullo_07}.
\\

\begin{figure*}[htbp]
    \centering
    \includegraphics[width=0.9\textwidth]{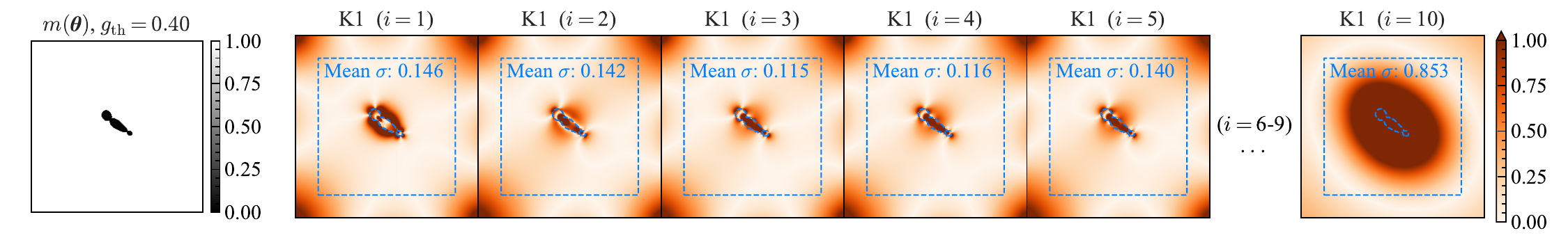}\\
    \includegraphics[width=0.9\textwidth]{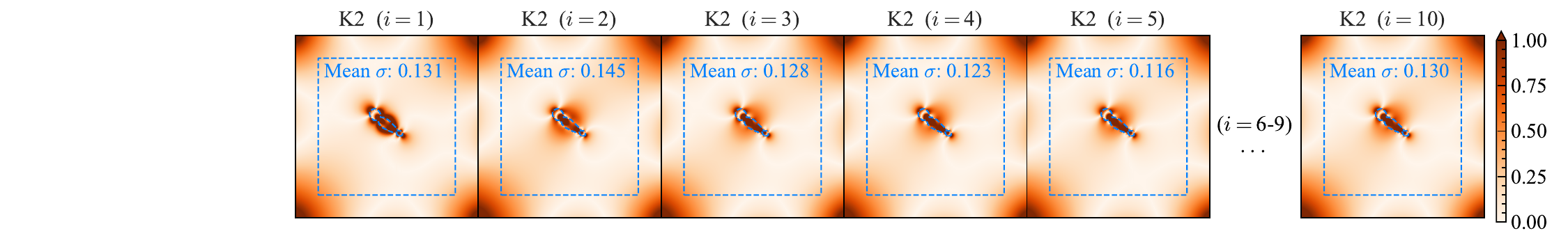}\\
    \includegraphics[width=0.9\textwidth]{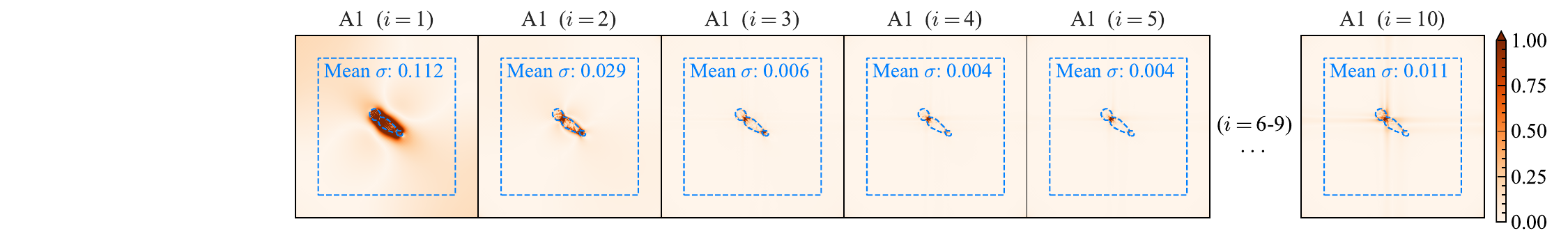}\\ 
    \includegraphics[width=0.9\textwidth]{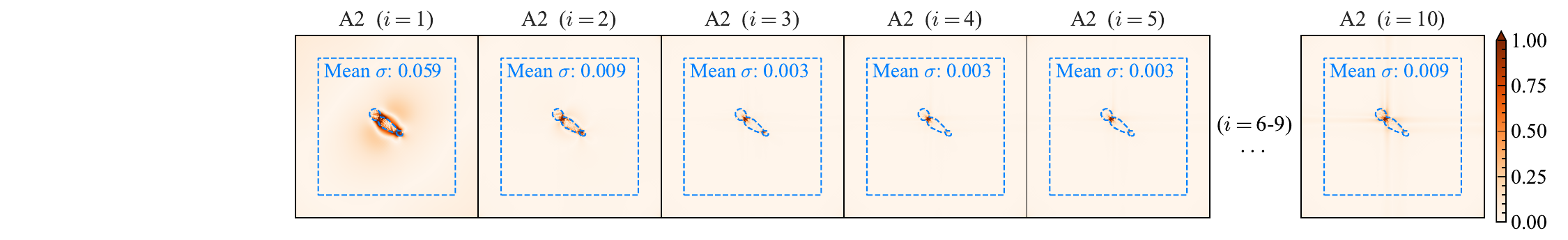}
    \caption{Normalized residual maps at iterations 1--5 and 10 for the toy model. Rows from top to bottom correspond to methods K1, K2, A1, and A2. Mean $\sigma$ values within the evaluation region are indicated in each panel.}    
    \label{fig:Toy_model_iter}
\end{figure*}

\begin{figure*}[htbp]
    \centering
    \includegraphics[width=0.9\textwidth]{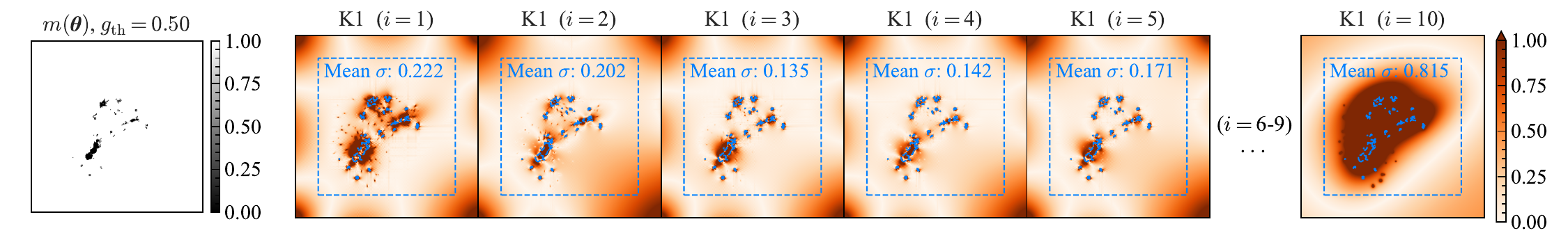}\\
    \includegraphics[width=0.9\textwidth]{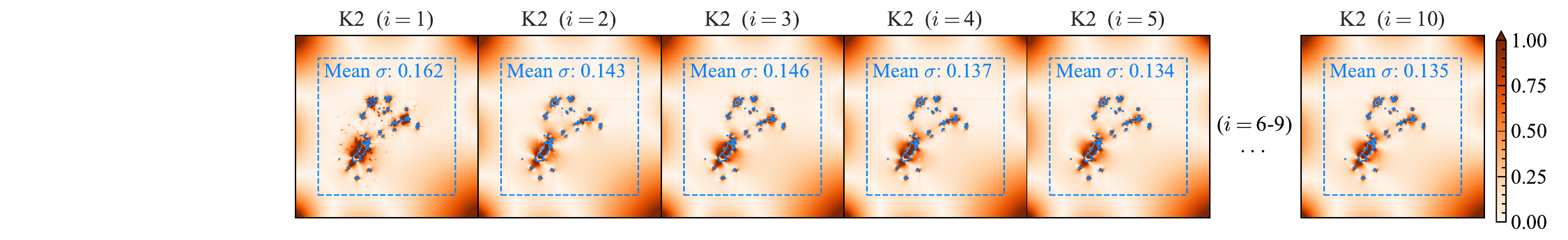}\\
    \includegraphics[width=0.9\textwidth]{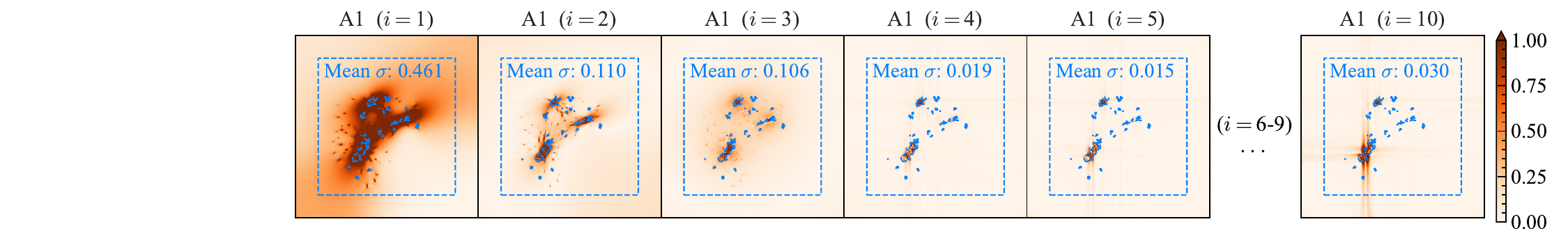}\\ 
    \includegraphics[width=0.9\textwidth]{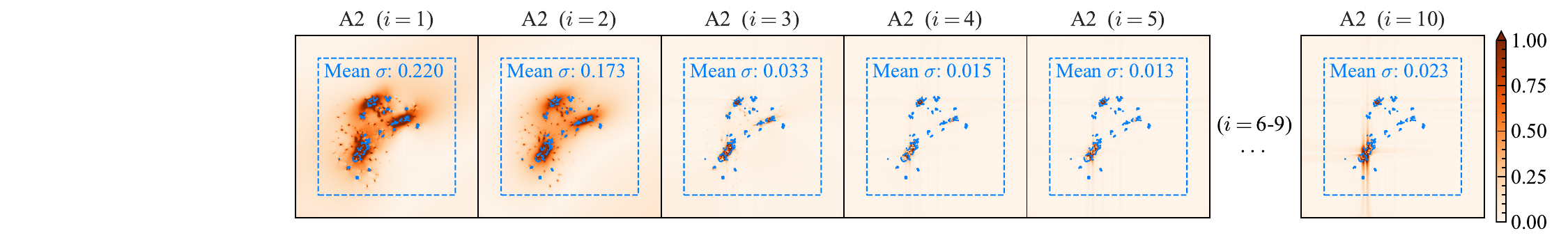}
    \caption{Normalized residual maps at iterations 1--5 and 10 for the B23 model. The panel layout and labeling follow the same convention as in Fig.~\ref{fig:Toy_model_iter}.}
    \label{fig:B23_iter}
\end{figure*} 

\appendix
\setcounter{section}{0}
\renewcommand{\thesection}{\Alph{section}}

\section{Detailed Iteration analysis} \label{app:iterations}
In weak lensing mass reconstruction of galaxy clusters, the convergence map is typically recovered from reduced shear through an iterative scheme, which ideally converges progressively toward the true solution. However, numerical instabilities can arise during the iteration, causing residuals to first decrease and then increase as iterations proceed.

To investigate this, we run each method for 10 iterations and present the results in Fig.~\ref{fig:Toy_model_iter} and Fig.~\ref{fig:B23_iter}. We find that different methods reach their minimum residuals at different iteration numbers. Remarkably, K1 diverges after reaching its lowest residuals in the B23 model cases. We also find that using a model-based initial guess improves iteration stability, suppressing the growth of residuals at later iterations.

Furthermore, in the final iterations of A1 and A2 methods, we observe mode leakage arising from numerical instability in matrix inversion in the AKRA-based methods. This effect is alleviated by adopting a smooth mask, which reduces spectral leakage and mode mixing caused by sharp mask boundaries \cite{Grain2009}, as shown by comparing Fig.~\ref{fig:A23_iter} and Fig.~\ref{fig:B23_iter}.

\section{The Mask Convolution Matrix} \label{app:mask}
The mask matrix arises from the convolution theorem, where multiplication in real space corresponds to convolution in Fourier space. The observed (masked) shear field is $\gamma_i^M(\vec{\theta}) = m(\vec{\theta})\gamma_i(\vec{\theta})$, which in Fourier space becomes
\begin{align}
\tilde{\gamma}_{i}^m(\vec{L}) 
&= \int \gamma_{i}(\vec{\theta})\, m(\vec{\theta})\, e^{-i \vec{L} \cdot \vec{\theta}}\, d^{2}\theta \nonumber \\
&= \int \frac{d^{2}\ell_{1}}{(2\pi)^{2}} \int d^{2}\ell_{2}\; 
    \tilde{\gamma}_{i}(\vec{\ell}_{1})\, \tilde{m}(\vec{\ell}_{2})\, 
    \delta^{D}(\vec{\ell}_{1}+\vec{\ell}_{2}-\vec{L}) \nonumber \\
&\propto \int d^{2}\vec{\ell}_{1}\; 
    \tilde{\gamma}_{i}(\vec{\ell}_{1})\, \tilde{m}(\vec{L}-\vec{\ell}_{1}).
\label{eq:gamma_convolution}
\end{align}
$\tilde{\gamma}_i(\vec{\ell}_1)$ and $\tilde{m}(\vec{\ell}_2)$ are the Fourier transforms of the true shear and mask functions, and its shape is $N_{\ell} \times N_{\ell}$, where $N_{\ell}$ is the total number of Fourier modes.
The discrete form of this convolution defines the mask matrix: $M({\vec{L},\vec{\ell}_1}) = \tilde{m}(\vec{L} - \vec{\ell}_1)$.

In practice, $\mathbf{M}$ is constructed column by column. For each Fourier mode $\vec{\ell}_1$: (1) Insert a delta function at mode $\vec{\ell}_1$ in Fourier space, (2) Inverse Fourier transform to real space to obtain a complex exponential $e^{i\vec{\ell}_1\cdot\vec{\theta}}$, (3) Multiply by the mask $m(\vec{\theta})e^{i\vec{\ell}_1\cdot\vec{\theta}}$, (4) Transform back to the Fourier space. This gives the column of $\mathbf{M}$ for mode $\vec{\ell}_1$. 
The resulting matrix has dimension $N_\ell^2 \times N_\ell^2$, and is generally a dense matrix. 

There are two important properties of $\mathbf{M}$ follow directly from its construction: \begin{itemize} 
\item \textit{Mask-free case:} In the case where $m(\vec{\theta}) = 1$ for all $\vec{\theta}$, its Fourier transform is $\tilde{m}(\vec{L}) = \delta^D(\vec{L})$, this yields $\mathbf{M} = \mathbf{I}$ and AKRA reduces to the standard KS inversion. 
\item \textit{Block-Toeplitz structure:} Since $M({\vec{L},\vec{\ell}_1})$ determined entirely by the mode difference $\vec{L} - \vec{\ell}_1$, the matrix has a block-Toeplitz structure, which makes its construction computationally efficient.
\end{itemize} 

The mask convolution matrix $\mathbf{M}$ encodes non-local couplings introduced by the mask, each Fourier mode of the observed shear contains contributions from neighboring modes of the true field. Reconstructing the full shear field thus requires solving an inverse problem that accounts for these mode couplings.

\bibliographystyle{apsrev}
\bibliography{ref}

\end{document}